\documentclass[runningheads]{llncs}
\usepackage[T1]{fontenc}
\usepackage{graphicx}
\begin{document}
\title{Machine Unlearning for Responsible and Adaptive AI in Education}
%
%
\author{Betty Mayeku\inst{1}\orcidID{0000-1111-2222-3333} \and
Sandra Hummel\inst{2}\orcidID{1111-2222-3333-4444} \and
Parisa Memarmoshrefi\inst{3}\orcidID{2222--3333-4444-5555}}
\authorrunning{B. Mayeku, S. Hummel and P. Memarmoshrefi}
%
\institute{Leipzig University, Germany 
\email{betty.mayeku@uni-leipzig.de}\\
 \and
Technical University Dresden, Germany\\
\email{sandra.hummel@tu-dresden.de}
\and
University of Göttingen, Germany\\
\email{memarmoshrefi@informatik.uni-goettingen.de}}

\maketitle              
\begin{abstract}
Machine Unlearning (MU) has emerged as a promising approach to addressing persistent challenges in Machine Learning (ML) systems. By enabling the selective removal of learned data, MU introduces protective, corrective, and adaptive capabilities that are central to advancing Responsible and Adaptive AI. However, despite its growing prominence in other domains, MU remains underexplored within education, a sector uniquely characterized by sensitive learner data, dynamic environments, and the high-stakes implications of algorithmic decision-making. This paper examines the potential of MU as both a mechanism for operationalizing Responsible AI principles and a foundation for Adaptive AI in ML-driven educational systems. Drawing on a structured review of 42 peer-reviewed studies, the paper analyzes key MU mechanisms and technical variants, and how they contribute to the practical realization of Responsible and Adaptive AI. Four core intervention domains where MU demonstrates significant promise are identified: privacy protection, resilience to adversarial or corrupted data, fairness through bias mitigation, and adaptability to evolving contexts. Furthermore, MU interventions are mapped to the technical, ethical, and pedagogical challenges inherent in educational AI. This mapping illustrates MU’s role as a strategic mechanism for enhancing compliance, reinforcing ethical safeguards, and supporting adaptability by ensuring that models remain flexible, maintainable, and contextually relevant over time. As a conceptual contribution, the paper introduces MU4RAAI, a reference architecture integrating MU within Responsible and Adaptive AI frameworks for educational contexts. MU is thus positioned not merely as a data deletion process but as a transformative approach for ensuring that educational AI systems remain  ethical, adaptive, and trustworthy.


\end{abstract}
\section{Introduction}
Artificial Intelligence (AI) has become deeply embedded in the institutional fabric of education, shaping assessment processes, feedback mechanisms, and data-informed decision-making. These systems promise scalable personalization, improved efficiency, and data-informed guidance \cite{Abdullah2022,Akinwalere2022,Dembe2024,Ersozlu2024,Khanipoor2024,PutraPratama2023,Razaulla2022,Singh2025,Yahayal2024}. However, alongside these opportunities, growing concerns have emerged regarding the risks and unintended consequences of integrating AI in the education domain. Delegation of pedagogical judgment to computational processes raises new challenges of accountability, fairness, transparency, and control, contributing to reduction of trust in AI systems in education \cite{Bentley2023,Dignum2021,Rodrguez2023,Holmes2021,Marcotte25,Nguyen2023,Yahayal2024}.

To harness the full potential of AI in education, researchers and policy makers have increasingly emphasized the importance of Responsible AI \cite{Bentley2023,Dignum2021,Rodrguez2023,Göllner2024,Marcotte25,Nguyen2023,Wang2020}. The frameworks developed by governing bodies including UNESCO, the European Commission (EU), and the Organization for Economic Co-operation and Development (OECD) articulate principles such as privacy, transparency, fairness, reliability, and accountability as essential foundations for Responsible AI \cite{EU2022,Floridi2020,Holmes2021,OECD2023,UNESCO2021}. These frameworks underscore the ethical imperative that AI in education should not only enhance personalization and efficiency, but also uphold learners’ rights, agency, and safety.

However, in practice, operationalizing Responsible AI in education remains a significant challenge. AI-driven educational applications rely heavily on machine learning (ML) models trained on vast volumes of educational data \cite{Holmes2021,Khanipoor2024,Luckin2016,Zhang2023}. Such data are highly sensitive, often involving minors; situational and dynamic, reflecting changing institutional and pedagogical contexts; and high-stakes, influencing critical decisions \cite{Akinwalere2022,Baker2019,BarreraCastro2025,DMello2020,Hummel2023,Hummel2025,Marcotte25,Rose2021,Yahayal2024}. Consequently, educational technologies must not only be accurate and efficient but also remain open to revision and reversibility, allowing for the correction or removal of outdated, contested, or ethically problematic information.

A fundamental limitation of conventional ML architectures, however, are inherently irreversible. Once data are internalized into model parameters, they continue to influence the outcomes even when they become outdated or invalid. This persistence conflicts with legal mandates such as the GDPR’s “Right to be Forgotten” \cite{GDPR2016} and with educational expectations for systems that are flexible, revisable, accountable and can uphold agency (both institutional and learner centered) over how past data inform algorithmic decisions. 

This tension raises a critical question: How can AI-driven educational systems be designed not only to learn responsibly, but also to selectively unlearn when data become outdated, contested, or pedagogically misaligned?

One promising approach that has potential to address this challenge is Machine Unlearning (MU). MU refers to a family of techniques that enable trained models to selectively remove the influence of specific data without requiring complete retraining \cite{Cao2015,Hine2024,Sai2024}. MU has been recognized for its potential to provide protective, corrective, and adaptive mechanisms \cite{Bourtoule2021,Cao2015,Gupta2021,Hine2024,Zhang2023} that enhance model reliability, accountability, and responsiveness to changing data and contexts. While these capabilities have attracted significant attention in domains such as healthcare and data governance, MU remains largely unexplored in education, despite the sector’s high sensitivity to data ethics, fairness, regulatory compliance and flexibility. 

This paper explores the potential of MU as a mechanism for advancing Responsible and Adaptive AI in education. Drawing on a structured review of 42 peer-reviewed studies, it identifies core MU intervention areas and introduces MU4RAAI, a conceptual reference architecture for integrating selective unlearning into educational ML workflows. The contribution is primarily conceptual, clarifying how MU can strengthen responsible and adaptive AI ecosystems in education.

The remainder of the paper is structured as follows. Section 2 provides an overview of ML applications in education and an overview of the MU concept. Section 3  highlights the methodology used in the study. In section 4, part 1 examines the distinctive characteristics and challenges of ML-driven education systems, and part 2 examines MU as a strategy for Responsible and Adaptive AI. Section 5 presents potential applications of MU in the education context by mapping MU interventions to the identified ML-driven education challenges with a focus on adaptability and Responsible AI. Section 6 introduces the MU4RAAI framework. Section 7 concludes with implications, limitations, and directions for future research.

\section{Machine Learning in Education and the Emergence of Machine Unlearning}
\subsection{Machine Learning in Education}
ML has transformed how learning is guided, assessed, and supported. From intelligent tutoring systems and adaptive learning platforms to predictive analytics, ML models underpin the personalization and automation of educational processes \cite{Ersozlu2024,Hummel2025,Kamalov2023,Khanipoor2024,PutraPratama2023}.

However, as highlighted in the introduction, ML models internalize patterns from vast amounts of learner data, making complete data removal non-trivial. Once trained, these models can retain the influence of sensitive or outdated data, even after they are removed from the databases \cite{Ali2024,Bourtoule2021,Sai2024,Singh2025,Zhang2023}. This persistence can lead to biased or inaccurate outcomes and complicates compliance with privacy regulations and ethical standards. Therefore, the need for mechanisms that allow ML models to completely “forget” has become increasingly urgent in educational contexts.

\subsection{Overview of Machine Unlearning}
MU, first formalized by \cite{Cao2015} and later refined by \cite{Ginart2019}, enables models to selectively forget or remove the influence of specific data from trained models without requiring full retraining. 

MU methods can target unlearning at various levels including instance-level, feature-level, and concept-level, depending on the nature of the data to be removed \cite{Ali2024,Blanco-Justicia2025,Chundawat2023,Nguyen2024,Sai2024}. According to \cite{Ali2024}, instance-level unlearning focuses on eliminating individual data points from a trained model. Feature-level unlearning suppresses or removes specific attributes or variables, especially when they are irrelevant, noisy, or biased. Concept-level unlearning involves selectively discarding  or updating knowledge linked to specific ideas or patterns. Its capacity to remove outdated or misleading knowledge makes it particularly valuable in dynamic and evolving contexts.
 
The concept of MU has attracted a lot of attention in fields such as computer vision, speech recognition, healthcare,finance \cite{Bourtoule2021,Li2025,Zhang2023}, where data privacy and model adaptability are critical. However, despite its promising potential, it remains underexplored in education. This gap is striking given the sector is defined by evolving data and context, ethical sensitivities, and the pedagogical imperative for reversibility. As such, the education landscape represents a uniquely urgent context for advancing research on MU.

\section{Methodology}
The goal of this work was to examine whether MU can respond meaningfully to the challenges of ML in education, particularly in relation to fostering Responsible and Adaptive AI. To interpret ML-driven education challenges normatively, this study adopted a conceptual lens grounded in five principles of Responsible AI that recur across global ethical frameworks, including the EU High-Level Expert Group on AI and UNESCO Recommendation on AI Ethics. These principles include (a) privacy, (b) security, (c) fairness and inclusion, (d) accountability and transparency, and (e) reliability and safety.

To inform this inquiry, a structured literature review was conducted in four major databases: IEEE Xplore, SpringerLink, Scopus, and Google Scholar. The search strategy combined keywords such as "AI in Education", “Application of machine learning in education”, “AI challenges in education”, “Machine unlearning applications”, and “Machine unlearning in education”. The inclusion criteria were articles from peer-reviewed journals or conferences published between 2015 and 2025, and focused on higher education contexts. Exclusion criteria included preprints, non-peer-reviewed contributions, and studies focusing solely on primary or secondary education.

After screening and deduplication, 42 publications were retained. Of these, 24 addressed challenges of ML in education, 18 focused on MU in general, and none directly examined MU within education-specific systems. This notable absence underscores the research gap that motivates the present study.

\section{Findings}
Publication trends demonstrate that interest in MU has increased notably in the 2023–2025 period, confirming its status as an emerging research domain. The absence of studies investigating MU in an educational context, while not evidence of irrelevance, indicates a critical research gap that motivates the present study. The following sections present detailed findings of the review.

\subsection{Characteristics and Challenges of ML in Education}
ML in education is shaped by a set of distinctive characteristics that enable its transformative potential and expose its inherent vulnerabilities. These characteristics distinguish educational ML applications from those in other domains. In particular, the sensitive nature of educational data, the contextual dependence and  variability of learning environments, and the complexity of pedagogical requirements introduce specific risks that can amplify broader systemic challenges. Understanding and addressing these challenges requires examining the defining characteristics of ML in education, the risks they entail, and how they collectively contribute to the broader landscape of ML-related challenges identified in the literature.

\subsubsection{Characteristics and Associated Risks.} ML-driven education systems share six distinct characteristics in that they rely on sensitive data, operate in dynamic and situational contexts, impact high-stakes decisions, continuously ingest new data streams and they are prone to bias.

\paragraph{Sensitive and Confidential Data:} Data is a fundamental component for ML-driven solutions in education \cite {Khanipoor2024}. The effectiveness of education platforms supporting personalised learning, learning analytics, adaptive systems, recommendation platforms, intelligent tutor among others, rely on collecting and analyzing vast amounts of learners' data. However, not only are educational data abundant, but learners’ data used in ML models often include highly sensitive and personal information such as  personal identifiers, grades, behavioral logs and even medical information \cite{Ali2024,Ersozlu2024,Sai2024,Yahayal2024} among others. Because learners are often minors, handling such data entails strong privacy obligations. Once trained, models can memorize this data, leaving learners vulnerable to privacy attacks \cite{Sai2024}. In addition, the use of surveillance-based data and the commercial exploitation of educational data raise additional concerns about the privacy of educational data \cite{Khanipoor2024,BarreraCastro2025,Marcotte25}. For these reasons, users may want a system to forget certain sensitive data and its entire lineage \cite{Ali2024,Nguyen2024,Zhang2023}. Therefore, it is paramount to ensure that learners' data is protected from unauthorized access and breaches, as well as comply with Regulations such as GDPR, which explicitly require mechanisms for the complete erasure of personal data upon request. However, most current ML systems cannot completely “forget” data once learned.

\paragraph{Situational and Context-specific Use:} Learning is inherently situated, occurring relative to the teaching environment and its pedagogical, institutional, and cultural context \cite{Illinoisun2012,Renkl2001}. Consequently, educational data extend beyond performance metrics to encompass refined indicators such as moments of confusion, hesitation, motivation, and doubt. While this richness enhances the potential of ML models to personalize and adapt, it also creates fragility. Data that are valid in one setting may become irrelevant or even harmful when applied in another. This situational dependence undermines the fairness, generalizability, and transferability of ML models, which are often optimized for specific curricula, institutional policies, or cultural assumptions \cite{BarreraCastro2025,Hummel2023}. As a result, deploying models across contexts without adaptation risks misinterpretation, biased outcomes, and diminished trust in AI-driven educational tools.

\paragraph{Dynamic and Evolving Environments:} The digital and education landscapes are often dynamic. Learners’ skills, progress, preferences, and needs evolve, while curricula, assessments, and regulations undergo continuous revision. Static ML models become quickly outdated, while noisy or incorrect entries degrade the model quality 
\cite{Chundawat2023,Nguyen2024}. ML models that fail to adapt in such dynamic environments risk fossilizing outdated assumptions, creating mismatches between what learners need and what systems deliver. Therefore, to maintain model performance and keep it current, it is necessary to remove obsolete or irrelevant knowledge and integrate new data.

\paragraph{High-stakes Outcomes:} ML-driven decisions in education can influence grades, career pathways, access to scholarships, targeted interventions, among others. However, data are often accompanied by noise or incorrect entries, which can seriously degrade the quality of the outcomes \cite{Chundawat2023,Nguyen2023}. For example: predictive models that identify “at-risk” learners or allocate resources can shape long-term opportunities and academic pathways, a single misclassified diagnostic entry (due to e.g.,noise, malicious data, out-of-distribution data) or mislabeled learner profile can propagate through predictive models and recommendation systems, locking students into inappropriate or inequitable academic pathways. Such errors or biases in these systems may not only carry life-changing consequences, reinforcing inequities, but also result in less user trust \cite{Nguyen2023,Nguyen2024}. There is therefore a need to clean up unnecessary data or correct mistakes to provide a better experience and accurate outcomes of the applications.

\paragraph{Continuous Data Collection:} Modern educational platforms rely on real-time monitoring of Learnng Management Systems (LMS) click streams, tutoring interactions, and online assessments \cite{Abdullah2022,Ali2024}. This produces vast, heterogeneous datasets of uneven quality. In addition, continuous data streams risk the ingestion of unauthorized or poisoned data. This makes educational systems vulnerable to data poisoning and corruption attacks, which compromise the integrity and accuracy of the models \cite{Nguyen2024,Sai2024}. Security breaches can propagate harmful behaviors across models. This situation is made more difficult by the way educational platforms are built. Data flows across learning management systems, institutional servers, and commercial analytics tools. Oversight is fragmented and safeguards across institutions and vendors remain inconsistent \cite{BarreraCastro2025,Hummel2023}. Hence, the need for a mechanism that allows removal of harmful data or mistakes after training.

\paragraph{Risk of Bias and Inequity:} The biased nature of ML algorithms is often a consequence of the underlying biases in the data \cite{Sai2024}. ML systems trained on educational data often inherit biases from language, gender, or socioeconomic patterns, among others.  If educational training data is biased or unrepresentative, these biases can become deeply ingrained, leading to discriminatory predictions and inequitable outcomes \cite{Ali2024,BarreraCastro2025,Marcotte25,Sai2024}. If biased results are used to drive decisions about learners and educational systems,  for example, mislabel learners based on incorrect algorithms or biased data \cite{Marcotte25}, the effects could be carried forth from individual to system levels. For instance, efficiency-designated recommendation systems can narrow learning pathways, overlooking diversity in preferences and backgrounds. Adaptive platforms may also erode learner autonomy if temporary behaviors are misinterpreted as fixed traits, creating feedback loops that further disadvantage some learners \cite{Hummel2025}. Without corrective mechanisms, these potential harms could outweigh, or at least challenge, the benefits of using ML in education, leading to social disparities and unfairness or reinforcing educational inequalities instead of reducing them \cite{Ali2024,BarreraCastro2025,Marcotte25}. 

\subsubsection{Linking Characteristics to Challenges.}
\begin{figure}
\includegraphics[width=\textwidth]{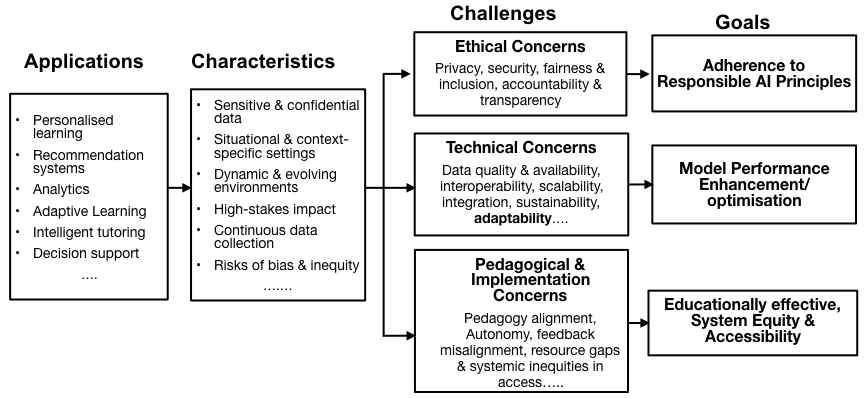}
\caption{Linking Characteristics to Challenges} \label{Fig1}
\end{figure}

These characteristics, i.e, sensitivity of data, situational dependence, dynamism, high-stakes impact, continuous data flows, and systemic bias, intersect in ways that generate overlapping vulnerabilities that undermine both the effectiveness and trustworthiness of educational AI systems. As illustrated in Fig.~\ref{Fig1}, these vulnerabilities converge into three recurrent clusters of challenges consistently highlighted in the literature. They include (a) ethical (b) technical and (c) pedagogical \& implementation challenges \cite{Abdullah2022,BarreraCastro2025,Dembe2024,Hummel2025,Khanipoor2024,Razaulla2022,Yahayal2024}.

\paragraph{Ethical challenges} involve concerns about privacy, security, fairness, accountability, and transparency. These concerns influence the operationalization of the core principles of Responsible AI.

\paragraph{Technical challenges} include issues with data quality and availability, interoperability, scalability, and adaptability to evolving environments. These challenges affect long-term optimization and performance of the model.

\paragraph{Pedagogical \& implementation challenges} encompass diminished learner autonomy, misaligned feedback, insufficient educator preparedness, inequities in access, and limited institutional capacity for integration. Such challenges influence the pedagogical soundness, inclusivity, and educational value of ML-driven systems.

The defining characteristics of ML-driven educational systems amplify these systemic challenges. For instance, the sensitivity of educational data heightens ethical risks concerning privacy and security. Situational dependence and dynamism expose technical limitations in adaptability, scalability, and interoperability while simultaneously raising ethical concerns about fairness and inclusivity. High-stakes decision-making amplifies both ethical and pedagogical concerns around accountability and learner agency. Continuous data collection intensifies challenges related to oversight, data drift, and model reliability. Systemic bias reinforces the issues of fairness and transparency across all layers of the system. 

Understanding this linkage is critical for identifying targeted interventions that mitigate ethical and technical vulnerabilities while preserving educational soundness.

\subsubsection{The Problem of Irreversibility.} 
Beyond these vulnerabilities, perhaps the most fundamental limitation of current ML systems lies in their irreversibility. Once data has been internalized, few mechanisms exist to remove its influence, even when the data is erroneous, outdated, or for which consent has been withdrawn. Thus, the trustworthiness of ML in education depends not only on accuracy but also on the ability of these systems to revise, retract, and forget data when necessary. In educational contexts, this need extends beyond compliance with data protection laws such as the GDPR. It speaks to a deeper pedagogical imperative - to empower learners and educators with meaningful agency over how their data shape algorithmic outcomes and to promote systems that are corrective, revisable, and adaptive rather than static and self-reinforcing. Addressing this problem requires exploring how reversibility mechanisms can be integrated into educational ML applications in ways that are technically sound and pedagogically meaningful.

\subsection{Machine Unlearning as a Strategy for Responsible and Adaptive AI}

One approach with significant potential to address these challenges is MU. Although it remains an emerging research field, its transformative potential in overcoming the persistent limitations of conventional ML systems has been increasingly recognized in recent studies \cite{Hine2024,Sai2024,Xu2023,Zhang2023}. 
This section examines key MU mechanisms, technical variants, and how they can provide a foundation for embedding Responsible AI principles while enhancing the adaptability and long-term resilience of AI systems.

\subsubsection{Machine Unlearning Mechanisms.} By design, MU has been recognized in the literature as a means to strengthen three core dimensions of AI reliability,  namely \emph{compliance}, \emph{correction}, and \emph{adaptability}. These dimensions are operationalized through several key MU mechanisms commonly discussed in recent studies. They include:

\paragraph{Data Minimization:} ML models often retain influence from all data ever used in training, including sensitive, outdated, or irrelevant data \cite{Ali2024,Bourtoule2021,Cao2015,Hine2024,Li2025,Zhang2023}. MU therefore provides a mechanism to “forget” specific data points or subsets without retraining from scratch, in accordance with the principle that only the data necessary for a task should be retained.

\paragraph{Post-Hoc Correction:} This refers to fixing mistakes after a model has already been trained \cite{Goel2024,Li2025}. If wrong or biased data enter the training set, retraining the whole model is costly. MU therefore allows the targeted removal of erroneous or harmful data after training, avoiding costly full retraining cycles when mistakes, bias, or poisoned data are discovered \cite{Goel2024,Li2025}.

\paragraph{Long-Term Model Maintenance and Adaptability:} 
Institutional and regulatory environments evolve continuously. As a result, older data may become invalid, obsolete, or non-compliant. Conventional ML systems are inherently cumulative, often retaining outdated information and biases. MU offers a mechanism for sustaining model relevance and trustworthiness over time by selectively removing obsolete knowledge and adapting to new data \cite{Gupta2021,Hine2024,Li2025,Sai2024,Zhang2023}. 

\subsubsection{Machine Unlearning Approaches and Techniques.}
The aforementioned mechanisms are operationalized through several technical variants of MU. Understanding these variants is essential for assessing their suitability for educational applications, which are uniquely constrained by regulatory requirements, institutional infrastructures, and pedagogical objectives.

As MU remains an emerging and rapidly evolving field, its conceptual and technical taxonomies often exhibit overlap, reflecting the diversity of mechanisms, application contexts, and levels of theoretical assurance observed in the literature \cite{Bourtoule2021,Hine2024,Sai2024,Xu2023,Zhang2023}. The absence of a unified taxonomy arises because MU techniques may differ not only in their operational mechanisms such as retraining, gradient manipulation, or knowledge transfer but also in their underlying objectives, ranging from strict data erasure to approximate or context-specific unlearning. Consequently, MU methods are best understood as existing along a continuum rather than as mutually exclusive categories.

To provide conceptual clarity and a basis for systematic comparison, this paper consolidates and organizes the diverse MU techniques reported in recent literature into five major approaches. These approaches are distinguished by their underlying principles, computational efficiency, implementation complexity, and strength of unlearning guarantees \cite{Li2025,Liu2025,Sai2024,Wang2024,Xu2023,Zhang2023}. They include:

\paragraph{Exact Unlearning:} is a full retraining-based approaches that ensure complete data removal by retraining from scratch, offering the strongest guarantees of compliance and reversibility \cite{Li2025,Sai2024,Wang2024,Xu2023,Zhang2023}.  However, it is computationally intensive and impractical for large-scale or continuously learning systems. Representative methods include Full Retraining and the Sharded, Isolated, Sliced, and Aggregated (SISA) training framework \cite{Bourtoule2021,Xu2023}.

\paragraph{Approximate Unlearning:} refers to techniques that aim to efficiently approximate the effect of full retraining  without actually retraining the entire model from scratch. These methods prioritize efficiency by making analytical or procedural adjustments to the model parameters or learned representations\cite{Gupta2021,Ginart2019,Li2025,Sai2024,Wang2024,Xu2023,Zhang2023}.They do not guarantee perfect removal but they improve computational efficiency, scalability and are practical for large models. Common techniques used include influence-function–based updates, certified unlearning, and gradient or parameter adjustment.

\paragraph{Gradient Reversal and Parameter Update:} is a specialized subclass of parameter-based unlearning where opposite gradients are applied to reverse the contribution of specific samples. It counteracts data influence by applying negative gradients \cite{Geng2025,Graves2020}. These methods are efficient for continuously learning systems and can be implemented incrementally. However, may lead to instability and lacks theoretical guarantees. Methods applied include reverse gradient descent (Anti-Gradient Updates), forgetting through adversarial optimization and selective layer unlearning.

\paragraph{Knowledge Distillation–Based Unlearning:} uses a teacher–student architecture, where the student model learns from the teacher’s outputs, excluding forgotten data. The student thus inherits generalizable knowledge without the influence of removed samples.\cite{Geng2025,Graves2020}. This technique allows for the preservation of model utility and performance. It is also scalable, and adaptable for continuous learning contexts. However, it introduces approximation errors and depends heavily on the quality of the distillation process and temperature tuning. Some of the techniques this approach uses include selective distillation \cite{Chundawat2023}, amnesiac distillation \cite{Graves2020}, among others.

\paragraph{Federated and Distributed Unlearning:} extends unlearning concepts to distributed and federated settings where data and models are decentralized. They enable local data deletion and aggregation consistency across distributed nodes\cite{Wang2024,Xu2023,Zhang2023}. These methods address the additional challenges of data privacy, and partial model aggregation. However, they face challenges related to communication overhead and ensuring verifiable unlearning consistency across distributed infrastructures. Representative approaches include client update removal and secure aggregation unlearning \cite{Gao2022,Liu2023}. Some hybrid models combine parameter adjustment or distillation mechanisms within distributed architectures, further blurring categorical boundaries.

These approaches represent a comprehensive synthesis of the evolving MU landscape thereby providing a structured foundation for analyzing how MU contributes to the practical realization of Responsible and Adaptive AI.

\subsubsection{Aligning Machine Unlearning with Responsible and Adaptive AI.} As mentioned earlier, MU strengthens three central dimensions of AI reliability, namely, compliance, correction, and adaptability. Through these dimensions, MU serves as a bridge between ethical principles and technical optimization, enabling AI systems to remain both responsible and adaptive over time. These relationships are elaborated below and summarized in Table~\ref{tab:MU_alignment}.

\begin{table}
\small
\centering
\caption{Alignment of MU Mechanisms and Approaches with Responsible and Adaptive AI Dimensions}
\label{tab:MU_alignment}
\begin{tabular}{|p{2.8cm}|p{3.1cm}|p{3.1cm}|p{2.8cm}|}
\hline
\textbf{MU Mechanism} & \textbf{Supporting Techniques} & \textbf{Primary Outcome} & \textbf{Responsible \& Adaptive AI Dimension} \\
\hline
Data Minimization & Exact Unlearning, Federated Unlearning & Compliance with data protection and erasure mandates & Privacy Protection, Institutional Accountability \\
\hline
Post-Hoc Correction & Approximate Unlearning, Gradient Reversal \& Parameter Update & Removal of biased, poisoned, or unauthorized data & Fairness, Security, Reliability \\
\hline
Long-Term Model Maintenance \& Adaptability & Knowledge Distillation–Based, Federated Unlearning & Sustained relevance and responsiveness of models to evolving contexts & Adaptability, Performance Optimization \\
\hline
\end{tabular}
\end{table}

Different MU mechanisms contribute to distinct but complementary aspects of Responsible and Adaptive AI.  \emph{Data Minimization} supports \emph{privacy protection} and \emph{regulatory compliance} by ensuring that models do not retain unnecessary or revoked information. \emph{Post-hoc Correction} enhances \emph{fairness}, \emph{security}, and \emph{reliability} by enabling the removal of corrupted, biased, or adversarially poisoned data from trained models without full retraining. \emph{Long-Term Model Maintenance and Adaptability} sustains \emph{model relevance} by facilitating selective unlearning of outdated or misaligned data, preventing fossilized errors and enabling dynamic adjustment to evolving contexts.

The implementation of these mechanisms depends on specific MU approaches and techniques.  
\emph{Exact Unlearning} (full retraining-based removal) provides the strongest compliance guarantees, ensuring verifiable data deletion but at high computational cost.  
\emph{Approximate Unlearning} (including certified unlearning and parameter adjustment) balances efficiency with reliability, supporting responsive correction and fairness repair.  
\emph{Gradient Reversal and Parameter Update} techniques enable local corrections of biased or poisoned gradients, reinforcing model robustness and accountability.  
\emph{Knowledge Distillation–Based Unlearning} transfers retained knowledge while discarding targeted information, maintaining accuracy and generalization even after selective forgetting.  
Finally, \emph{Federated and Distributed Unlearning} aligns with privacy-by-design principles by supporting decentralized, institution-level control over data and model updates—crucial for distributed educational infrastructures.

Together, these mechanisms and approaches operationalize Responsible AI by embedding ethical safeguards (privacy, fairness, accountability) and advance Adaptive AI by ensuring models remain flexible, maintainable, and contextually relevant over time. 

\section{Machine Unlearning Interventions in Educational AI}
Building upon the preceding discussion on embedding MU within Responsible and Adaptive AI frameworks, this section contextualizes how MU can be operationalized as an intervention strategy in educational ML-driven systems. While MU has gained traction across various domains such as healthcare, cybersecurity, and finance, its application within the education sector remains largely unexplored. This absence is notable given the sector’s growing reliance on ML-driven analytics, the sensitivity of learners’ data, and the high-stakes implications of algorithmic decisions in adaptive learning, assessment, and policy contexts.

Traditionally, educational AI research has emphasized preventive mechanisms such as privacy-by-design and explainability. However, less attention has been devoted to post-hoc strategies that allow selective data removal or model correction once learning has occurred. MU offers a complementary mechanism—enabling not only compliance with data protection and ethical mandates but also restoring learner agency, pedagogical integrity, and contextual relevance as educational environments evolve.

Accordingly, this section outlines the potential technical implications of MU for educational AI systems (Section 5.1) and maps MU interventions to the core challenges of ML-driven education (Section 5.2). Together, these discussions illustrate how MU can advance Responsible and Adaptive AI in education by aligning ethical imperatives with computational design.

\subsection{Technical Implications of MU in Educational AI Systems}
The technical characteristics of MU methods have distinct implications for the design, governance, and pedagogical viability of AI systems in education. Each unlearning approach presents trade-offs between computational efficiency, compliance assurance, adaptability, and educational alignment. For example: \emph{Exact Unlearning } ensures compliance with legal frameworks. However, its computational cost and retraining demands render it impractical for educational platforms that continuously collect data from learner interaction. This makes it more suitable for small-scale or batch-updated institutional models, such as admissions or assessment analytics, than live learning platforms. \emph{Approximate and Certified Unlearning} offers agility for dynamic pedagogical contexts e.g., updating learner models or risk predictions but requires transparent oversight to prevent residual bias or privacy risks.

\emph{Gradient-based unlearning and Parameter Adjustment} aligns with adaptive and continuously learning systems, such as personalized recommendation engines or intelligent tutoring systems, enabling real-time correction, although it demands stability safeguards to maintain pedagogical coherence.\emph{Knowledge Distillation} supports lifelong refinement in curriculum-aligned models but requires validation to ensure forgotten data are not implicitly retained. \emph{Federated and Distributed Unlearning} is essential for cross-platform educational ecosystems, ensuring that deletion requests propagate across institutions and systems while preserving privacy 

In practice, the educational sector, where both technical and ethical considerations intersect, hybrid strategies combining  federated architectures with approximate unlearning and institutional audit layers may provide the most pragmatic pathway toward realizing Responsible and Adaptive AI ecosystem.

\subsection{Mapping MU Interventions to ML-Driven Education Challenges}
\begin{figure}
\includegraphics[width=\textwidth]{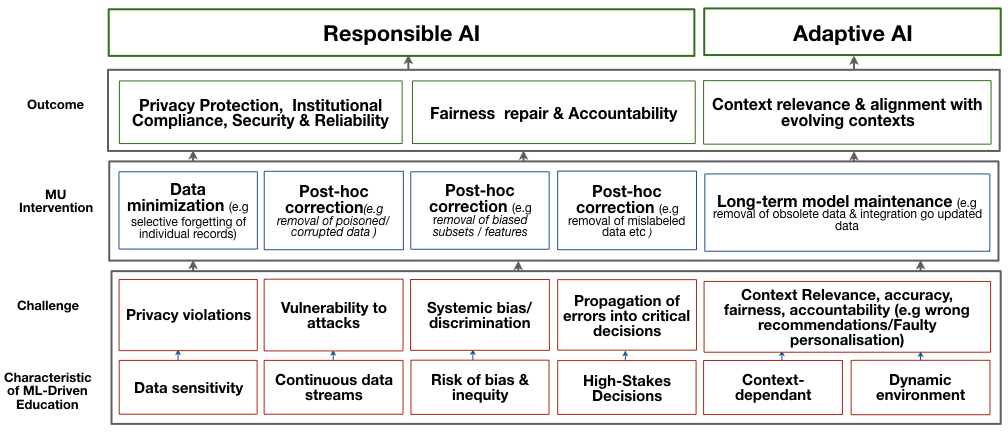}
\caption{Mapping MU Interventions to ML-Driven Education Challenges} \label{Fig2}
\end{figure}

As discussed earlier, ML-driven educational systems are characterized by distinctive features (data sensitivity, situational dependence, dynamism, high-stakes decision impact, continuous data collection, and systemic bias). These characteristics give rise to intertwined technical, ethical, and pedagogical challenges. MU provides a structured intervention pathway to mitigate these challenges by enabling selective forgetting, model correction, and contextual adaptation as illustrated in Fig.~\ref{Fig2} and further elaborated in the following.

\subsubsection{Privacy Protection (Sensitive Data):} Educational ML models often process highly sensitive learner information. MU supports privacy protection through data minimization and selective influence removal, ensuring that deleted data no longer contribute to model behavior\cite{Ali2024,Nguyen2024,Sai2024,Zhang2023}. In this way, MU extends compliance beyond storage-level deletion to model-level forgetting, aligning with data protection mandates such as the GDPR and enhancing institutional data governance.

\subsubsection{Security and Reliability (Continuous Data Flows):} Given that educational systems continuously ingest real-time interaction data, they are vulnerable to corrupted or adversarial inputs \cite{Chundawat2023,Nguyen2024,Sai2024}. MU’s post-hoc correction mechanisms allow targeted removal of poisoned data or compromised features, restoring model integrity and ensuring resilience against data poisoning or drift.

\subsubsection{Bias Mitigation and Fairness (High-Stakes Decisions):} ML models in education can perpetuate systemic inequities or reinforce misclassifications with significant learner consequences. MU enables post-hoc fairness repair through selective unlearning of biased instances or mislabeled samples. By removing their influence, MU restores equity in predictions and recommendations while reinforcing accountability and trustworthiness.

\subsubsection{Adaptability and Contextual Relevance (Context dependency and Dynamic Environments):} 
Education is inherently dynamic, with evolving institutional demands, pedagogical objectives, and learner contexts. MU supports adaptability by unlearning outdated or pedagogically irrelevant data, enabling continuous model recalibration. This ensures that AI systems remain contextually relevant, ethically aligned, and pedagogically responsive.

These interventions demonstrate that MU extends beyond a reactive data deletion mechanism but it functions as a proactive framework that aligns ethical governance with technical optimization while enhancing the resilience and contextual relevance of educational ML systems.

\section{Machine Unlearning for Responsible and Adaptive AI (MU4RAAI) Framework}
The mappings presented in the preceding section form the conceptual foundation of the Machine Unlearning for Responsible and Adaptive AI (MU4RAAI) framework in Fig.~\ref{Fig3}, which synthesizes MU intervention pathways into an integrated model for educational contexts. As illustrated in Fig.~\ref{Fig3}, the framework positions unlearning as both an ethical safeguard and a technical enabler. It captures the dynamic interaction between ML-driven educational systems, their associated challenges, MU interventions, and the resulting Responsible and Adaptive AI outcomes. The MU4RAAI framework is organized into four interrelated layers:

\begin{figure}
\includegraphics[width=\textwidth]{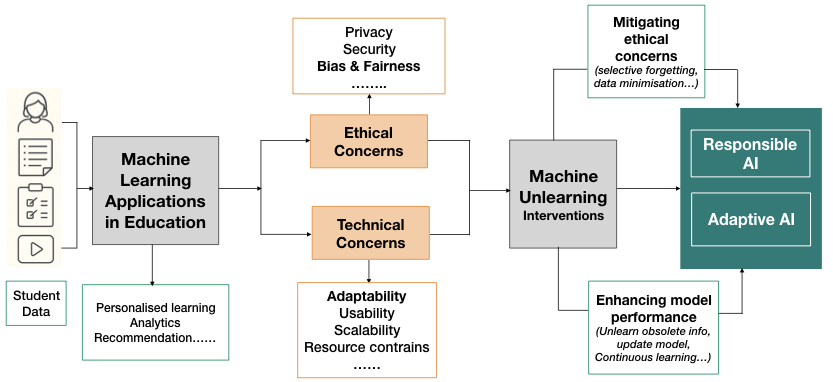}
\caption{Machine Unlearning for Responsible and Adaptable AI (MU4RAAI) Framework in Educational Context} \label{Fig3}
\end{figure}

\paragraph{Input Layer (ML-driven Education Systems):} Represents existing educational AI systems such as predictive analytics, adaptive learning platforms, recommendation systems, etc., that depend on educational data. These systems exhibit the defining characteristics discussed earlier.

\paragraph{Challenge Layer:} Highlights the ethical, technical, and pedagogical vulnerabilities that arise from these system characteristics, including privacy and security breaches, bias propagation, limited adaptability etc.

\paragraph{MU Intervention Layer:} Compasses the core unlearning interventions for mitigating specific vulnerabilities through selective forgetting, correction, or/and continuous refinement of model knowledge.

\paragraph{Outcome Layer (Responsible and Adaptive AI):} Demonstrates the dual impact of MU interventions. On one hand, they strengthen Responsible AI by ensuring privacy, security and fairness. On the other hand, they promote Adaptive AI by enhancing contextual relevance, responsiveness, and long-term performance stability.

The framework depicts how MU interventions operate across different layers of educational AI systems to address the ethical and technical vulnerabilities identified earlier. Ethically, MU operationalizes Responsible AI principles such as privacy, security and fairness by enabling models to selectively forget, correct, or revise their learned representations. This translation from abstract ethical ideals into concrete model behaviors ensures that educational AI systems are not only compliant with regulatory standards but also aligned with human-centered values and educational equity. Technically, MU strengthens Adaptive AI capabilities by supporting post-hoc correction and continuous model evolution. It enhances adaptability and long-term model reliability through the removal of corrupted, outdated, or irrelevant data thereby preventing the accumulation of “fossilized” errors and maintaining system relevance in changing educational contexts. This positions MU as a strategy for Adaptive AI, ensuring that models remain dynamic, context-sensitive, and performant over time. 

By integrating these dual pathways, MU4RAAI provides a conceptual bridge between normative AI ethics frameworks (e.g., UNESCO, OECD, and EU AI Guidelines) and the operational realities of educational ML systems. By integrating unlearning into the model life cycle, the framework advances a vision of AI that is both ethically grounded and technically adaptive, capable of learning responsibly, correcting itself, and evolving alongside the educational environments it serves.

\section{Conclusion}
This paper explored the potential of MU to address persistent challenges in ML applications within educational contexts. Drawing from prior literature, four core application areas were identified: privacy protection, security and robustness, bias mitigation for fairness, and adaptability with performance optimization. These applications are enabled by the ability of MU to selectively remove data influence from trained models, providing a practical mechanism for data minimization, post-hoc correction, and sustainable model maintenance without full retraining. These capacities are particularly relevant in educational settings that exhibit distinctive ethical and operational requirements amplified by its unique characteristics including sensitivity of data, evolving and situation specific context, and governance constraints. Beyond its technical utility, MU serves as a strategic enabler for aligning legal compliance frameworks, ethical design principles, and performance standards in AI-driven education. Advancing its application can pave the way for more responsive, equitable, and trustworthy learning technologies that embody both Responsible and Adaptive AI principles. This study contributes to this direction through the proposed MU4RAAI framework, which provides a conceptual foundation for integrating MU into responsible and adaptive educational AI systems. Future research should focus on refining MU techniques for these specific conditions and empirically evaluate their pedagogical, ethical, and technical effectiveness within real-world educational ecosystems.
\begin{credits}
\subsubsection{\ackname} The authors acknowledge the financial support by the Federal Ministry of Research, Technology and Space of Germany and by Sächsische Staatsministerium für Wissenschaft, Kultur und Tourismus in the programme Center of Excellence for AI-research „Center for Scalable Data Analytics and Artificial Intelligence Dresden/Leipzig", project identification number: ScaDS.AI

\end{credits}


\end{document}